\documentclass[graybox]{svmult}

\usepackage{graphicx}
\usepackage{epstopdf}
\usepackage{mathptmx}       % selects Times Roman as basic font
\usepackage{helvet}         % selects Helvetica as sans-serif font
\usepackage{courier}        % selects Courier as typewriter font
\usepackage{type1cm}        % activate if the above 3 fonts are
\usepackage{makeidx}         % allows index generation
\usepackage{graphicx}        % standard LaTeX graphics tool
\usepackage{multicol}        % used for the two-column index
\usepackage[bottom]{footmisc}% places footnotes at page bottom
\usepackage[margin=1in]{geometry}

\usepackage{latexsym,epsfig,graphicx,subfigure,amsmath,amssymb,amscd,multirow,psfrag,paralist,dsfont,float}
\makeindex             % used for the subject index
\newcommand{\thetavec}{{\boldsymbol{\theta}}}

\newcommand{\avar}{{\rm AVar}}

\newcommand{\xvec}{\boldsymbol{x}}

\makeatletter
\renewcommand\@biblabel[1]{#1.}

\makeatother

\begin{document}

\title*{Bayesian Sequential Design Based on Dual Objectives for Accelerated Life Tests}
\titlerunning{Sequential Design for Accelerated Life Tests}
\author{Lu Lu, I-Chen Lee, and Yili Hong}
\authorrunning{Lu, Lee, and Hong}
\institute{Lu Lu \at Department of Mathematics \& Statistics, University of South Florida, Tampa, FL, \email{lulu1@usf.edu}
\and I-Chen Lee \at Department of Statistics, National Cheng Kung University, Tainan, Taiwan, \email{iclee@mail.ncku.edu.tw}
\and Yili Hong,  corresponding author \at Department of Statistics, Virginia Tech, Blacksburg, VA 24061, \email{yilihong@vt.edu}}

%%%%%%%%%%%%%%%%%%%%%%%%%%%%%%%%%%%%%%%%%%%%%%%%%%%%%%%%%%%%%%%%%%%%%%%%%%%%%%%%%%%%%%%%%
\maketitle
%%%%%%%%%%%%%%%%%%%%%%%%%%%%%%%%%%%%%%%%%%%%%%%%%%%%%%%%%%%%%%%%%%%%%%%%%%%%%%%%%%%%%%%%%
%online version abstract
\abstract*{Traditional accelerated life test plans are typically based on optimizing the C-optimality for minimizing the variance of an interested quantile of the lifetime distribution. The traditional methods rely on some specified planning values for the model parameters, which are usually unknown prior to the actual tests. The ambiguity of the specified parameters can lead to suboptimal designs for optimizing the intended reliability performance. In this paper, we propose a sequential design strategy for life test plans based on considering dual objectives. In the early stage of the sequential experiment, we suggest to allocate more design locations based on optimizing the D-optimality to quickly gain precision in the estimated model parameters. In the later stage of the experiment, we can allocate more samples based on optimizing the C-optimality to maximize the precision of the estimated quantile of the lifetime distribution. We compare the proposed sequential design strategy with existing test plans considering only a single criterion and illustrate the new method with an example on fatigue testing of polymer composites.}

%%%%%%%%%%%%%%%%%%%%%%%%%%%%%%%%%%%%%%%%%%%%%%%%%%%%%%%%%%%%%%%%%%%%%%%%%%%%%%%%%%%%%%%%%

%print version
\abstract{Traditional accelerated life test plans are typically based on optimizing the C-optimality for minimizing the variance of an interested quantile of the lifetime distribution. The traditional methods rely on some specified planning values for the model parameters, which are usually unknown prior to the actual tests. The ambiguity of the specified parameters can lead to suboptimal designs for optimizing the intended reliability performance. In this paper, we propose a sequential design strategy for life test plans based on considering dual objectives. In the early stage of the sequential experiment, we suggest to allocate more design locations based on optimizing the D-optimality to quickly gain precision in the estimated model parameters. In the later stage of the experiment, we can allocate more samples based on optimizing the C-optimality to maximize the precision of the estimated quantile of the lifetime distribution. We compare the proposed sequential design strategy with existing test plans considering only a single criterion and illustrate the new method with an example on fatigue testing of polymer composites.}


%\tableofcontents
%\newpage

%%%%%%%%%%%%%%%%%%%%%%%%%%%%%%%%%%%%%%%%%%%%%%%%%%%%%%%%%%%%%%%%%%%%%%%%%%%%%%%%%%%%%%%%%
\section{Introduction}\label{sec:introuduction}
%%%%%%%%%%%%%%%%%%%%%%%%%%%%%%%%%%%%%%%%%%%%%%%%%%%%%%%%%%%%%%%%%%%%%%%%%%%%%%%%%%%%%%%%%
\subsection{Background}
%%%%%%%%%%%%%%%%%%%%%%%%%%%%%%%%%%%%%%%%%%%%%%%%%%%%%%%%%%%%%%%%%%%%%%%%%%%%%%%%%%%%%%%%%
For many long life products or systems, accelerated lifetime tests (ALTs) \cite{MeekerEscobar1998,EscobarMeeker2006} are broadly used to accelerate the failure process by exposing the units under harsher conditions than usual and to collect timely information for effectively predicting the lifetime under the normal operating conditions. An ALT plan is often chosen to minimize the anticipated variance of estimated reliability metric under an interested use condition given an assumed ALT model.

This paper considers an example on planning fatigue tests for polymer composites. Polymer composites are popularly used in many fields of industry such as aircraft, wind turbine, transportation, construction and manufacture, because of their many desirable features such as light weight, high strength, and long-term durability. However, the performance of polymer composites can change after long periods of use due to the fatigue of the materials resulted from being exposed to varied stress levels. Hence fatigue tests of polymer composites aim to assess the material's reliability at some specified normal stress levels. Due to the extreme long life property of polymer composites, the ALTs are desired to acquire failure information at much higher stress levels than the normal use conditions.

In a typical ALT setting, $n$ units are tested under some elevated test conditions. Let $x_i$ represent the stress level at which unit $i$ is tested, and $t_i$ represents the recorded failure time (e.g., the number of cycles for the polymer composite fatigue test) if a failure was observed during the test or the censoring time if unit $i$ had not failed by the end of the test. Then an ALT model is fitted to relate the failure time with the stress level, which is used to predict the product reliability at other stress levels under normal use conditions. A particular interesting problem for planning an ALT is to determine the stress levels at which individual units will be tested given an affordable sample size. In this case, the number of test units are already determined based on the available resources and any time or budget constraints. The goal of the test plan is to choose the stress levels, $x_i$ for $i=1,\cdots, n$, that offer the best precision of the estimated reliability of interest.

ALT plans have been studied extensively in the past few decades. Most of the methods focus on finding the optimal test plans that maximize some specified utility functions based on the available information. Since the ALT failure models are typically non-linear, the associated information matrices usually depend on the model parameters, and hence the selected optimal designs are dependent on the values of the model parameters used at the planning stage. Given the true values of the model parameters are usually not known precisely at the planning stage, the resulted ALT plans could be suboptimal for assessing the interested reliability quantities. Bayesian methods have been utilized to leverage prior knowledge on the model parameters through carefully specified prior distributions, which allow the uncertainty of the prior information to be properly propagated through the statistical inference and offer more flexibility in combining information from various sources.

In addition, the special challenges for polymer composites tests also involve the use of expensive test equipment and the lack of prior information on the performance of new composites materials. Due to the limited test equipment (it is not uncommon that each test laboratory may only have a couple of testing machines available), simultaneously testing multiple samples are usually impractical. As a result, the tests need to be performed sequentially, which offers an opportunity to improve our limited prior knowledge on the model parameters as more data are collected. In this case, it is natural to employ a Bayesian sequential design to select the next design point based on optimizing the expected utility function over the prior distribution based on the current information of the model parameters updated by the observed data, which is captured by the posterior distribution conditioned on the observed failure times from the tested units.

Given the different choices of the utility functions for measuring different aspects of test performance, the sequential Bayesian design strategy can result in different test plans. For example, the commonly used D-optimality criterion focuses on achieving the most precise estimation of the model parameters of an ALT model by maximizing the determinant of the information matrix. While the C-optimality criterion aims to maximize the precision of a linear function of the model parameters by minimizing the its asymptotic variance. In the ALT setting, the C-optimal designs are often used for obtaining the most precise prediction of some interested quantile of the lifetime distribution under the normal use conditions, which involves extrapolating the stress variable outside the range of observed values (at the elevated stress levels). The optimal designs based on different objectives will push in different directions for allocating test units at different stress levels. Hence, an optimal test plan based on a single criterion may select suboptimal design when other criteria are also of primary interest. This has motivated us to consider multiple objectives in the design selection process.

In this paper, we consider dual objectives in the sequential Bayesian design setting. More specifically at the early stage of the sequential experiment, the sequential runs will be selected based on considering the  D-optimality to quickly gain most precision in the estimated model parameters. Then in the later stage, more sequential runs will be selected based on the C-optimality to gain most precision of the predicted quantile of the lifetime distribution at the normal use conditions. It is expected that by considering the dual objectives at different stages of the sequential experiment, we can seek for more balanced performance of design on both the estimation and prediction. In addition, improving the estimation of the model parameters at the early stage of the sequential experiment when little prior information was available can improve our estimation of the anticipated variance of the interested quantile of the lifetime distribution, and hence result in selecting a more effective test plan for optimizing product reliability at the normal use conditions.

%%%%%%%%%%%%%%%%%%%%%%%%%%%%%%%%%%%%%%%%%%%%%%%%%%%%%%%%%%%%%%%%%%%%%%%%%%%%%%%%%%%%%%%%%
\subsection{Related Literature}
%%%%%%%%%%%%%%%%%%%%%%%%%%%%%%%%%%%%%%%%%%%%%%%%%%%%%%%%%%%%%%%%%%%%%%%%%%%%%%%%%%%%%%%%%

Traditional non-Bayesian methods for designing an ALT are based on properties of maximum likelihood (ML) estimators. Meeker and Escobar \cite{MeekerEscobar1998} provided a general guideline for planning life tests to obtain the precise prediction at the normal use condition. Several work such as Chernoff~\cite{chernoff1962} and Meeker and Hahn \cite{meeker1977,meeker1985} studied optimum designs and compromise plans and outlined practical guidelines for planning an efficient ALT. Some recent developments on ALT include Ye et al. \cite{YeHongXie2013}, Pan and Yang \cite{PanYang2014}, Tsai et al. \cite{TsaiJiangLio2015}, and Ng et al. \cite{Ngetal2017}. Escobar and Meeker~\cite{EscobarMeeker2006} provided a review on the accelerated test models. Limon et al. \cite{LimonYadavLiao2017} provided a review on planning and analysis of the accelerated tests for assessing reliability.

Bayesian techniques have been broadly used in design of experiments. Bayesian methods assume prior distributions for the unknown parameters and the inference of unknown quantities are based on the posterior distribution of the parameters based on the observed data.  Chaloner and Verdinelli \cite{ChalonerVerdinelli1995} provided a comprehensive review of Bayesian experimental design techniques for linear and non-linear models, among which Bayesian D-optimal designs and Bayesian C-optimal designs are popular choices that have been used broadly in reliability test plans. Zhang and Meeker \cite{ZhangMeeker2006} developed a Bayesian test plan based on minimizing the pre-posterior expectation of the posterior variance over the marginal distribution of all possible test data. Hong et al. \cite{Hongetal2015} proposed new Bayesian criteria based on the large-sample approximations, which offer simplified solutions to the Bayesian test plans.

Sequential test plans have been popular for reliability tests that are either very expensive or very time-consuming. Instead of determining the design locations (i.e., the stress levels) for all test units prior to the experiment, the test in a sequential design is determined and implemented one at a time given the current information gained from the observed data. At each step, the next optimal design point is determined by optimizing a design criterion summarized over the posterior distribution of the parameters given all the data observed so far. It was first introduced in the non-Bayesian framework using the ML estimator with the D- or C-optimality criterion for designing nonlinear experiments \cite{Wu1985,McLeishTosh1990}. In the Bayesian framework, sequential designs based on considering the D-optimality criterion were used often. For example, Dror and Steinberg \cite{DrorSteinberg2008} developed the Bayesian sequential D-optimal designs for generalized linear models. Roy et al. \cite{Royetal2009} and Zhu et al. \cite{Zhuetal2014} proved convergence properties of the Bayesian sequential D-optimal designs for different forms of models.

In planning for the fatigue tests for polymer composites, King et al.~\cite{king2016} proposed an optimum test plan for the constant amplitude cyclic fatigue testing of polymer composites material. Lee et al.~\cite{Leeetal2018} proposed a sequential Bayesian C-optimal test plan for the polymer composites fatigue testing, which selected the sequential design points based on optimizing the posterior asymptotic variance of an interested quantile across a range of normal use conditions.

%%%%%%%%%%%%%%%%%%%%%%%%%%%%%%%%%%%%%%%%%%%%%%%%%%%%%%%%%%%%%%%%%%%%%%%%%%%%%%%%%%%%%%%%%
\subsection{Overview}
%%%%%%%%%%%%%%%%%%%%%%%%%%%%%%%%%%%%%%%%%%%%%%%%%%%%%%%%%%%%%%%%%%%%%%%%%%%%%%%%%%%%%%%%%
The remaining of the paper is organized as follows. Section~\ref{sec:ALT} provides the basic background on the ALT plans. Section~\ref{sec:SBD} discusses the Bayesian sequential design which helps improve the efficiency of the fatigue testing plan when imprecise prior knowledge on the planning parameter values is available. Section~\ref{sec:dual.obj} proposes a new method based on considering dual objectives in Bayesian sequential design, where D-optimality is employed initially to quickly improve the precision of the estimated model parameters in the early runs and followed with C-optimal sequential runs to further improve the precision of the estimated reliability quantile at the specified normal use conditions. The new method will be illustrated in Section~\ref{sec:example} using the polymer composites fatigue testing example, and compared with the Bayesian sequential designs that consider only a single criterion. In the end, Section~\ref{sec:conclusion} offers additional discussion and conclusions.

%%%%%%%%%%%%%%%%%%%%%%%%%%%%%%%%%%%%%%%%%%%%%%%%%%%%%%%%%%%%%%%%%%%%%%%%%%%%%%%%%%%%%%%%%
\section{Accelerated Life Test Plans}\label{sec:ALT}
%%%%%%%%%%%%%%%%%%%%%%%%%%%%%%%%%%%%%%%%%%%%%%%%%%%%%%%%%%%%%%%%%%%%%%%%%%%%%%%%%%%%%%%%%
In this section, we give an introduction to the general ALT plans. We focus on the ALT plan with a single accelerating factor such as in the fatigue testing example for polymer composites. The ALTs involve testing units at different stress levels of the accelerating factor for quickly obtaining failure information for the product. Hence, the ALT plans require the determination of the levels of the accelerating factor to be implemented in the tests (i.e., the stress levels) and the sample size at each level (how many units to be tested at each stress level). Similar to the regular design of experiments, where optimal designs (e.g., D-, A-, G-, or I-optimality \cite{Montgomery2017}) are often chosen for achieving the best precision of some interested quantities (e.g., model parameters or predictions throughout the design space) under an assumed response model, the ALT plans are often based on optimizing the precision of some interested reliability metric, such as a quantile of the lifetime, given an assumed ALT model. Because the data is collected under the elevated stress levels to make inference on reliability at the normal use conditions, extrapolation on the accelerating factors are naturally involved in ALT plans and hence requires strong assumptions on the specified ALT models.

For modeling, the log-location-scale family of distributions (e.g., the Weibull and Lognormal) are often used to model the accelerated lifetime. The models often assume a common scale parameter but allowing the location parameter to change at the different stress levels through a parametric model. An optimal ALT plan is often determined based on minimizing the asymptotic variance of a reliability metric under the assumed model, which is dependent on the model parameters (e.g., the scale and location parameters) that are unknown prior to the data collection. Hence, the efficiency of the ALT plan is largely dependent on the choice of the parameter values at the planning stage.

For the polymer composites fatigue test example, let $T$ denote the cycles to failure which is assumed to follow a log-location-scale family of distribution with the cumulative distribution function (cdf) and the probability density function (pdf) given in the forms below
\begin{eqnarray*}
F(t;\boldmath{\theta})=\Phi\left[\frac{\log(t)-\mu}{\nu}\right], \quad \textrm{ and }\quad f(t;\boldmath{\theta})=\frac{1}{\nu t}\phi\left[\frac{\log(t)-\mu}{\nu}\right].
\end{eqnarray*}
In the equations above, $\mu$ and $\nu$ are the location and scale parameters, respectively. The $\Phi(\cdot)$ and $\phi(\cdot)$ are the cdf and pdf of the standard normal distribution. Let $\boldsymbol{\theta}=(\mu,\nu)$ denote the vector of the unknown parameters included in the model. The scale parameter $\nu$ is often assumed to be constant over the different stress levels, while the location parameter $\mu=\mu_\boldmath{\beta}(x)$ is assumed to be dependent on the stress level $x$ through model parameters $\boldmath{\beta}$. A physically motivated nonlinear model \cite{Epaarachchi2003} is used to model the relationship between the cycles-to-failure and the stress level as given in the form
\begin{equation}\label{eq:ECm}
\mu_\beta(x)=\frac{1}{B}\log\left\{\left(\frac{B}{A}\right)h^B\left(\frac{\sigma_{ult}}{x}-1\right)
\left(\frac{\sigma_{ult}}{x}\right)^{\gamma\left(\alpha\right)-1}\left[1-\psi\left(R\right)\right]^{-\gamma(\alpha)}+1\right\}.
\end{equation}
In Eqn. (\ref{eq:ECm}), the unknown parameters are $\beta=\left(A,B\right)$, where $A$ is a model parameter representing the environmental effect on the material fatigue, and $B$ is the material specific effect. The remaining involved variables that are known for the test planning include: the stress ratio, $R=\sigma_m/\sigma_M$, where $\sigma_M$ and $\sigma_m$ are the maximum and minimum stresses, the frequency of the cyclic stress testing, $h$, the ultimate stress of the material, $\sigma_{ult}$, and the smallest angle between the testing direction and the fiber direction, $\alpha$. In addition, functions of known parameters $\psi(R)$ and $\gamma(\alpha)$ are defined as
\begin{equation*}
\psi(R)=\left\{\begin{array}{ll}
R & \mbox{for } \infty<R<1 \\
\frac{1}{R} & \mbox{for } 1<R<\infty
\end{array}\right.\,\,,
\end{equation*}
and $\gamma(\alpha)=1.6-\psi|\sin(\alpha)|$, respectively. The empirical model in Eqn.  (\ref{eq:ECm}) is flexible to be applied to a wide variety of materials with different settings of stress levels, angles and frequencies. Then, in the assumed ALT model, the unknown parameters are $\thetavec = \left(A, B, \nu \right)$.

\begin{figure}
\centering
\includegraphics[width=.5\textwidth]{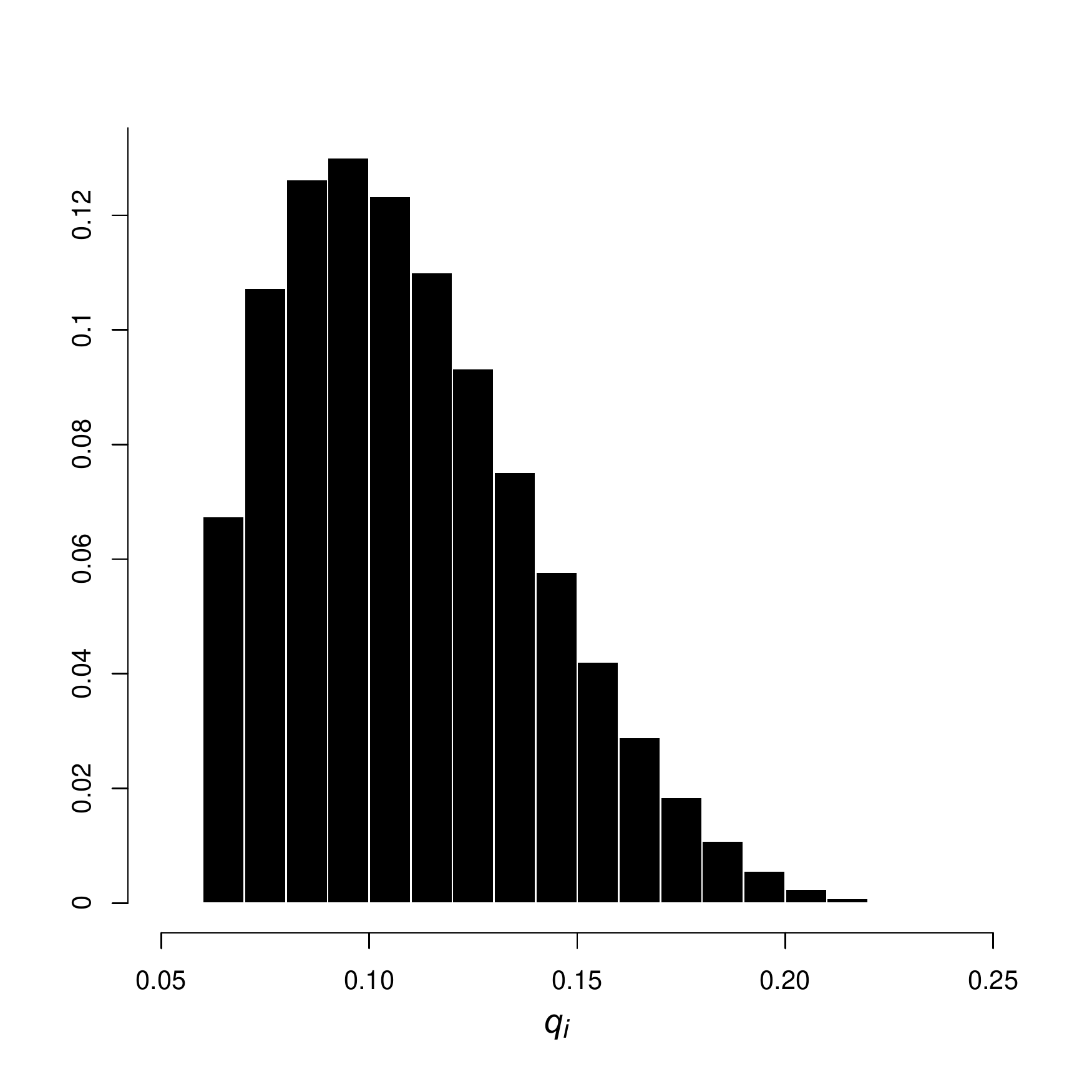}
\caption{An illustration of the use stress profile.}\label{fig:use.profile}
\end{figure}

In reliability analysis, the quantile of the lifetime in the lower tail of the distribution ($p\leq0.5$) is often of interest for capturing important reliability characteristics. Let $\zeta_{p,x}$ denote the $p$th quantile at the stress level $x$, which is related to the ALT model parameters through the following form
\begin{equation}\label{eq:qlt}
\log(\zeta_{p,x})=\mu_\boldmath{\beta}(x)+z_p\nu,
\end{equation}
where $z_p$ is the $p$th quantile of the standard log-location-scale distribution, and $x$ is a specified stress level under the normal use condition. The $p$th quantile of the lifetime can be estimated by substituting the model parameters in Eqn. (\ref{eq:qlt}) by its estimators, $\hat{\beta}$ and $\hat{\nu}$. The ALT plan can be chosen based on minimizing the asymptotic variance of the estimated $p$th quantile life, denoted by $\avar\left[\log\left(\hat{\zeta}_p,x\right)\right]$.
Considering in many real applications the stress level can vary across a range of use conditions, a use stress profile shown in Figure~\ref{fig:use.profile} is considered for the polymer composites fatigue test plan. The use condition stress levels range between $x_L=0.05\sigma_{ult}$ to $x_U=0.25\sigma_{ult}$. Let $\{x_1,\cdots,x_k\}$ denote all the use stress levels with $x_i=q_i \sigma_{ult}$ for $q_i \in \left[0.05,0.25\right]$ and $\{w_1, \cdots,w_k\}$ denote their relative frequencies in the use profile with $\sum_{k=1}^{K}w_k=1$. The total weighted asymptotic variance at all use stress levels is defined as
\begin{equation}\label{Eq:wavar}
\sum_{k=1}^{K}w_k \avar\left[\log\left(\hat{\zeta}_{p,x_k}\right)\right],
\end{equation}
for capturing the overall asymptotic variance throughout all possible use conditions.

The ML estimation is often used to calculate the weighted total asymptotic variance given in Eqn. (\ref{Eq:wavar}). The ML approach estimates the parameters based on maximizing the likelihood function of the parameters given the observed data. For the polymer composites fatigue test, the failure time data were collected within a predetermined test duration, and some test units had not failed by the end of the test period, which were censored observations. Let $(x_i,t_i,\delta_i)$ represent the observed data for the $i$th test unit, where $\delta_i$ is the censoring indicator and $\delta_i=1$ if the $i$th unit is censored and $\delta_i=0$ otherwise, and $t_i$ is the failure time when $\delta_i=0$ and the censoring time when $\delta_i =1$. Given the observed test data $\left(\mathbf{x}_n,\mathbf{t}_n,\mathbf{\delta}_n\right)$, where $\mathbf{x}_n=(x_1,\cdots,x_n)'$, $\mathbf{t}_n=(t_1,\cdots,t_n)'$, and $\mathbf{\delta}_n=(\delta_1,\cdots,\delta_n)'$, the likelihood function is given by
\begin{equation*}
L\left(\boldmath{\theta}|\mathbf{x}_n,\mathbf{t}_n,\mathbf{\delta}_n\right)=\prod_{i=1}^{n}\left\{\frac{1}{\nu t_i}\phi\left[\frac{\log(t_i)-\mu_\beta\left(x_i\right)}{\nu} \right] \right\}^{(1-\delta_i)}\left\{1-\Phi\left[\frac{\log(t_i)-\mu_\beta\left(x_i\right)}{\nu} \right]  \right\}^{\delta_i}.
\end{equation*}
Let $z_i=\left[\log(t_i)-\mu_\beta\left(x_i\right)\right]/\nu$, then the log-likelihood function is expressed as
\begin{equation}
l\left(\boldmath{\theta}|\mathbf{x}_n,\mathbf{t}_n,\mathbf{\delta}_n\right)=\sum_{i=1}^{n}\left(1-\delta_i\right)\left[\log\phi\left(z_i\right)-\log\left(\nu\right)-\log\left(t_i\right)\right]+\delta_i\log\left[1-\Phi\left(z_i\right)\right].
\end{equation}
The ML estimate of $\boldmath{\theta}$ is the solution to the equation $\partial l\left(\boldmath{\theta}\right)/\partial \boldmath{\theta}=\mathbf{0}$ and the asymptotic variance of the ML estimator, $\hat{\boldmath{\theta}}$ is given by
\begin{equation}\label{Eq:mleavar}
\Sigma_{\boldmath{\theta}}=I^{-1}_{n}\left(\boldmath{\theta, \mathbf{x}_n}\right)=\left\{E\left[-\frac{\partial^2 l\left(\boldmath{\theta}|\mathbf{x}_n,\mathbf{t}_n,\mathbf{\delta}_n\right)}{\partial \boldmath{\theta} \partial \boldmath{\theta}'}\right]\right\}^{-1}.
\end{equation}
Then the total weighted asymptotic variance in Eqn. (\ref{Eq:wavar}) can be calculated as
\begin{equation}\label{Eq:wavar2}
\sum_{k=1}^{K}w_k \avar\left[\log\left(\hat{\zeta}_{p,x_k}\right)\right]=\sum_{k=1}^{K}w_k \mathbf{c}'_k\Sigma_{\boldmath{\theta}}\mathbf{c}_k,
\end{equation}
where $\mathbf{c}_k=\left[\partial \mu_{\boldmath{\beta}}\left(x_k\right)/\partial A, \partial \mu_\boldmath{\beta}\left(x_k\right)/\partial B, z_p\right]'$. Note that using the ML estimation when there is a small number of observed failures with a high censoring rate can result in unstable estimates. If the censoring time is random, then the expectation-maximization approach can be used to obtain more stable estimates of the model parameters (e.g., Park and Lee \cite{ParkLee2012}).

The C-optimal design \cite{ChalonerVerdinelli1995}, which minimizes the asymptotic variance of the estimated quantile of the lifetime distribution at a range of normal use conditions measured by Eqn. (\ref{Eq:wavar2}) can be selected as the best ALT plan for determining the number of test units at different stress levels. Note the test plan based on minimizing Eqn. (\ref{Eq:wavar2}) should be performed prior to the data collection, and hence heavily depends on the values of the parameters, which are often unknown or at least not known precisely at the planning stage. Therefore, when little prior information was available on the true parameter values, an alternative to the ML approach for estimating the model parameters was proposed by Lee et al. \cite{Leeetal2018} based on using a Bayesian sequential design, which selects the next single design point based on optimizing Eqn. (\ref{Eq:wavar2}) over the posterior distribution of the parameter $\boldmath{\theta}$ given the currently observed data.

%%%%%%%%%%%%%%%%%%%%%%%%%%%%%%%%%%%%%%%%%%%%%%%%%%%%%%%%%%%%%%%%%%%%%%%%%%%%%%%%%%%%%%%%%
\section{Sequential Bayesian Design}\label{sec:SBD}
%%%%%%%%%%%%%%%%%%%%%%%%%%%%%%%%%%%%%%%%%%%%%%%%%%%%%%%%%%%%%%%%%%%%%%%%%%%%%%%%%%%%%%%%%
The sequential design framework used in this paper is similar to Lee at al. \cite{Leeetal2018}. Here we provide a brief description of the general framework within the context of the fatigue test plan for the polymer composites. Given a use stress profile, the $(n+1)$th design point was selected based on minimizing the average posterior asymptotic variance over the posterior distribution of the model parameters given the first $n$ observations, which is calculated by:
\begin{equation}\label{Eq:Bcopt}
\phi\left(x_{n+1}\right)=\int_\Theta\left[\sum_{k=1}^{K}w_k \mathbf{c}'_k\Sigma_{\boldmath{\theta}}\left(x_{n+1}\right)\mathbf{c}_k\right]\pi\left(\boldmath{\theta}|\mathbf{x}_n,\mathbf{t}_n,
\boldmath{\delta}_n\right)d\boldmath{\theta},
\end{equation}
where $\Sigma_{\boldmath{\theta}}\left(x_{n+1}\right) = \left[I_{n+1}\left(\theta,x_{n+1}\right)\right]^{-1}=\left[I_{n}\left(\theta,\mathbf{x}_n\right)+I_1\left(\theta,x_{n+1}\right)\right]^{-1}$.
Note in Eqn. (\ref{Eq:Bcopt}), the posterior distribution updated by the first $n$ observations, $\pi\left(\boldmath{\theta}|\mathbf{x}_n,\mathbf{t}_n,\boldmath{\delta}_n\right)$, can be considered a prior distribution of parameters prior to selecting the $(n+1)$th design point. A Markov chain Monte Carlo (MCMC) method was developed in Lee et al. \cite{Leeetal2018} to approximate the posterior distribution $\pi\left(\boldmath{\theta}|\mathbf{x}_n,\mathbf{t}_n,\boldmath{\delta}_n\right)$, where $\boldmath{\theta}=\left(A,B,\nu\right)$, for the selection of sequential optimal design points. Uniform distributions were assumed for both the environmental and physical parameters $A$ and $B$, as in $A\sim \mbox{Uniform}\left(a_1,a_2\right)$ and $B\sim \mbox{Uniform}\left(b_1,b_2\right)$, and $a_1, a_2, b_1$ and $b_2$ are constants specified based on the anticipated ranges of the parameters $A$ and $B$. An inverse gamma distribution with the shape parameter $\kappa$ and the scale parameter $\gamma$ was used for $\nu^2$ as it is the conjugate prior for the lognormal distribution, which is one of the commonly used log-location-scale distributions in reliability analysis.

In Lee et al. \cite{Leeetal2018}, two algorithms were provided to evaluate the anticipated asymptotic variance in Eqn. (\ref{Eq:Bcopt}) and to select the optimal next design location in the sequential Bayesian design. At each iteration of the sequential Bayesian design, the two algorithms are used (i) to draw samples from the posterior distribution $\pi\left(\boldmath{\theta}|\mathbf{x}_n,\mathbf{t}_n,\boldmath{\delta}_n\right)$ and use the Monte Carlo integration for approximating Eqn. (\ref{Eq:Bcopt}), and (ii) to  determine the optimal design location that minimizes Eqn. (\ref{Eq:Bcopt}) over all possible design points.

To encourage broad applications, an R package named ``SeqBayesDesign" \cite{packageSeq} was developed to implement the sequential Bayesian design developed in Lee et al. \cite{Leeetal2018}. The package can be applied to determine the sequential Bayesian design for traditional ALTs and the constant amplitude fatigue test based on either the log-normal distribution or the Weibull distribution. Users can easily implement the method by providing the inputs on the historical test data, anticipated use condition levels, the prior information on the model parameters which can be specified in the form of the uniform or normal distribution, and the candidate design points. After specifying the input settings, the optimal design point for the next step will be calculated by the two algorithms.

%%%%%%%%%%%%%%%%%%%%%%%%%%%%%%%%%%%%%%%%%%%%%%%%%%%%%%%%%%%%%%%%%%%%%%%%%%%%%%%%%%%%%%%%%
\section{Sequential Design Under Dual Objectives}\label{sec:dual.obj}
%%%%%%%%%%%%%%%%%%%%%%%%%%%%%%%%%%%%%%%%%%%%%%%%%%%%%%%%%%%%%%%%%%%%%%%%%%%%%%%%%%%%%%%%%
In this section, we describe a new sequential design strategy based on dual objectives. We first describe the motivation for considering dual objectives. Typically, the optimal design based on a single criterion can sacrifice its performance on aspects that are not measured by the chosen criterion. For example, the D-optimal designs maximize the information gain measured by the determinant of the information matrix, $|I_{n}\left(\theta, \xvec_n\right)|$. These designs allocate the design points to obtain the most precise estimates of the model parameters based on the data collected at the accelerated stress levels, and may not offer sufficient efficiency when predicting reliability at the normal use conditions which requires extrapolation beyond the range of the observed data. On the other hand, the C-optimal design, which focuses on obtaining the most precise prediction at the specified normal use condition(s), may be suboptimal for precisely estimating the model parameters. In recent decades, it has become more desirable to consider multiple aspects for design selection. The desirability function approach by Derringer and Suich \cite{DerringerSuich1980} has been broadly used to choose an optimal design based on combining multiple criteria through a user-defined desirability function that heavily relies on subjective user choices on weighting, scaling and metric form prior to understanding the design performance and potential trade-offs. Lu et al. \cite{Luetal2011} introduced the Pareto front approach to design selection and optimization by intentionally separating the objective and subjective selection stages and allows to make informed decisions based on quantitatively understanding the trade-offs and robustness to different user priorities.

We consider dual objectives in the sequential Bayesian design context. Given the sequential designs are often used in scenarios with small test units and limited prior knowledge on the model parameters, we propose to use the D-optimality at the early stage of the sequential experiment for a quick improvement of the precision of estimated model parameters, and then followed with the C-optimality criterion for more effective improvement on the precision of the estimated quantile of the lifetime at the normal use conditions. Considering the selection of the sequential runs are dependent on the current knowledge of the model parameters (either based on the assumed planning values or from the observed test data), when the planning values or the estimated parameters from limited observations are far off from the true values, using the D-optimality for choosing sequential runs is expected to quickly improve the precision of the estimated parameter values. Then with sufficiently precise estimates of model parameters, we have more precise estimate of asymptotic variance of the reliability quantity of interest, and hence can more efficiently allocate additional runs for directly improving the precision of predictions at the normal use conditions. When there are essentially sufficient number of runs collected, the Bayesian sequential design based on considering dual objectives may offer similar performance as the regular optimal designs based on considering a single criterion. However, for situations when very limited number of tests could afford, the dual objective sequential test is expected to offer more robustness and balanced performance between the two objectives on improving both the estimation and prediction outside the test region.

More specifically, suppose we want to design an $N$-run sequential Bayesian test plan and we want to choose the first $N_1$ runs based on the Bayesian D-optimality and the next $N-N_1$ runs based on the Bayesian C-optimality. Then, among the first $N_1$ runs, the $(n+1)$th run at $x_{n+1}$ is selected based on maximizing the expected D-optimality over the posterior distribution of the model parameters based on the first  $n$ observed test units as given by
\begin{equation}
\psi\left(x_{n+1}\right)=\int_\Theta \log\left\{|I_{n+1}\left(\theta,x_{n+1}\right)|\right\}\pi\left(\boldmath{\theta}|\mathbf{x}_n,\mathbf{t}_n,
\boldmath{\delta}_n\right)d\boldmath{\theta},
\end{equation}
where $|I_{n+1}\left(\theta,x_{n+1}\right)|$ is the determinant of the information matrix based on the first $n+1$ runs which can be updated from the information matrix for the first $n$ runs, $I_{n}\left(\theta,x_{n}\right)$, by the information gained from the additional observation at $x_{n+1}$ as in
\begin{equation}
I_{n+1}\left(\theta,x_{n+1}\right)=I_{n}\left(\theta,\mathbf{x}_n\right)+I_1\left(\theta,x_{n+1}\right).
\end{equation}
Then the optimal design point for the $(n+1)$th run is selected by
\begin{equation}
x_{n+1}^*=\textrm{argmax}_{x_{n+1}\in\left[x_L,\,\,x_U\right]}\psi\left(x_{n+1}\right).
\end{equation}
Then from the $N_1+1$ run on, the remaining design locations will be selected based on minimizing the total weighted asymptotic variance of the interested quantile of the lifetime distribution given in Eqn. (\ref{Eq:wavar2}) by seeking the optimal design location at $(n+1)$th observation as
\begin{equation}
x_{n+1}^*=\textrm{argmin}_{x_{n+1}\in\left[x_L,\,\,x_U\right]}\phi\left(x_{n+1}\right),
\end{equation}
until all $N$ runs are selected and executed.

The implementation of the dual objective Bayesian sequential design using the R package ``SeqBayesDesign" \cite{packageSeq} is convenient. This package allows the users to generate the Bayesian sequential optimal design based on a user choice on the optimization criterion between the D-optimality and C-optimality. By changing the optimization criterion among the sequential runs, we can easily implement the dual objective Bayesian sequential design for any desired fractions of the D-optimal and C-optimal runs. In the next section, we will illustrate the proposed method using the polymer composites fatigue testing example and compare a few Bayesian sequential test plans with different fractions of D-optimal runs to the Bayesian sequential designs considering only a single criterion. The R code for implementing the proposed method for the polymer composites fatigue test example is available from the authors upon request.

%%%%%%%%%%%%%%%%%%%%%%%%%%%%%%%%%%%%%%%%%%%%%%%%%%%%%%%%%%%%%%%%%%%%%%%%%%%%%%%%%%%%%%%%%
\section{Application in Polymer Composites Fatigue Testing}\label{sec:example}
%%%%%%%%%%%%%%%%%%%%%%%%%%%%%%%%%%%%%%%%%%%%%%%%%%%%%%%%%%%%%%%%%%%%%%%%%%%%%%%%%%%%%%%%%
In this section, we illustrate the sequential Bayesian design considering dual objectives using the polymer composites fatigue testing example. The original data consisting of 14 observations from a fatigue testing experiment for glass fibers were summarized in Lee et al. \cite{Leeetal2018}. Among the 14 existing observations, there were 11 failures and 3 right-censored observations. The ML estimates of the model parameters based on the 14 observations are $\hat{A}=0.00157$, $\hat{B}=0.3188$ and $\hat{\nu}=0.7259$, which are used as the best available estimates of model parameters when evaluating the performance of the sequential dual objective Bayesian designs. Other variables that were set for the test plan include $h=2, R=0.1, \alpha=0$, and $\sigma_{ult}=1339.67$.

\begin{figure}
\centering
\includegraphics[width=.5\textwidth]{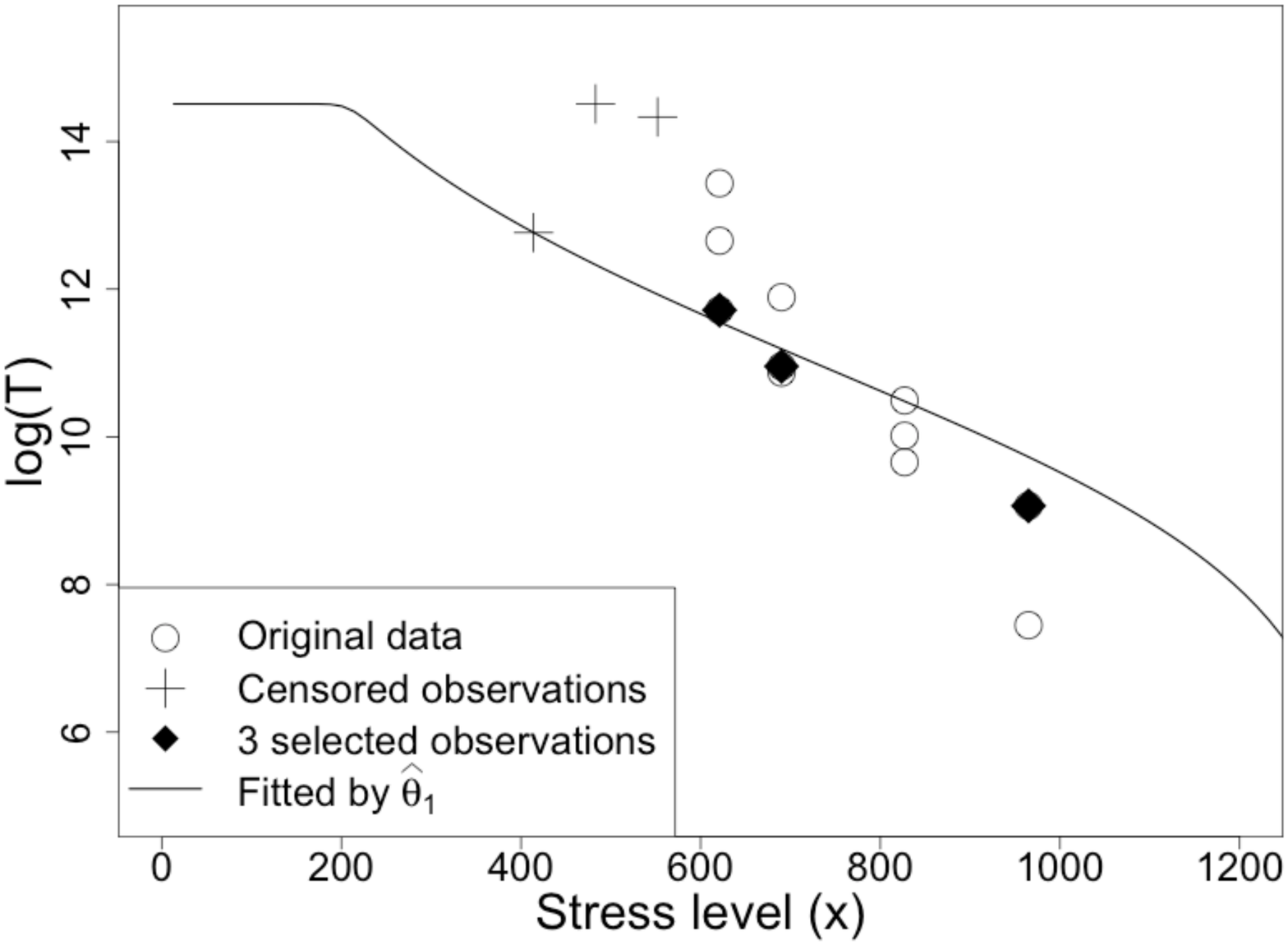}
\caption{Plot shows the three selected observations from the original data for Data Set 1 with its fitted stress-life relationship by $\hat{\thetavec}_1$.}\label{fig:DS1}
\end{figure}

Considering in many polymer composites fatigue testing, the available historical data are often limited, we select only three observations (the minimum number required to estimate all three model parameters) from the 14 observations to represent the limited existing data. First, we chose to evaluate the same subset of three observations that were evaluated in Lee et al. \cite{Leeetal2018}, namely the Data~Set~1, based on which the ML estimates of model parameters are $\hat{\thetavec}_1 = (0.0005, 0.7429, 0.1658)$. Figure~\ref{fig:DS1} shows the stress-life relationship for all the 14 observations and the subset of three that are considered as limited existing data in our example. We adapted the Bayesian sequential algorithms developed in Lee et al. \cite{Leeetal2018} for considering the dual objectives and implemented using the R package ``SeqBayesDesign" \cite{packageSeq}.

\begin{table}
\caption{Five sequential Bayesian designs for comparison.}\label{tab:fiveSBD}
\begin{center}
\begin{tabular}{l|l}\hline\hline
Design scenarios & Description \\ \hline
a: 12 C-opt & all 12 sequential runs are generated based on the Bayesian C-optimality criterion \\
b: 12 D-opt & all 12 sequential runs are generated based on the Bayesian D-optimality criterion \\
c: 6 D-opt $+$ 6 C-opt & first 6 runs are based on D-optimality and last 6 runs are based on C-optimality \\
d: 4 D-opt $+$ 8 C-opt & first 4 runs are based on D-optimality and last 8 runs are based on C-optimality \\
e: 2 D-opt $+$ 10 C-opt & first 2 runs are based on D-optimality and last 10 runs are based on C-optimality \\ \hline\hline
\end{tabular}
\end{center}
\end{table}

\vspace{0.5cm}

\begin{figure}
\begin{center}
\includegraphics[width=.5\textwidth]{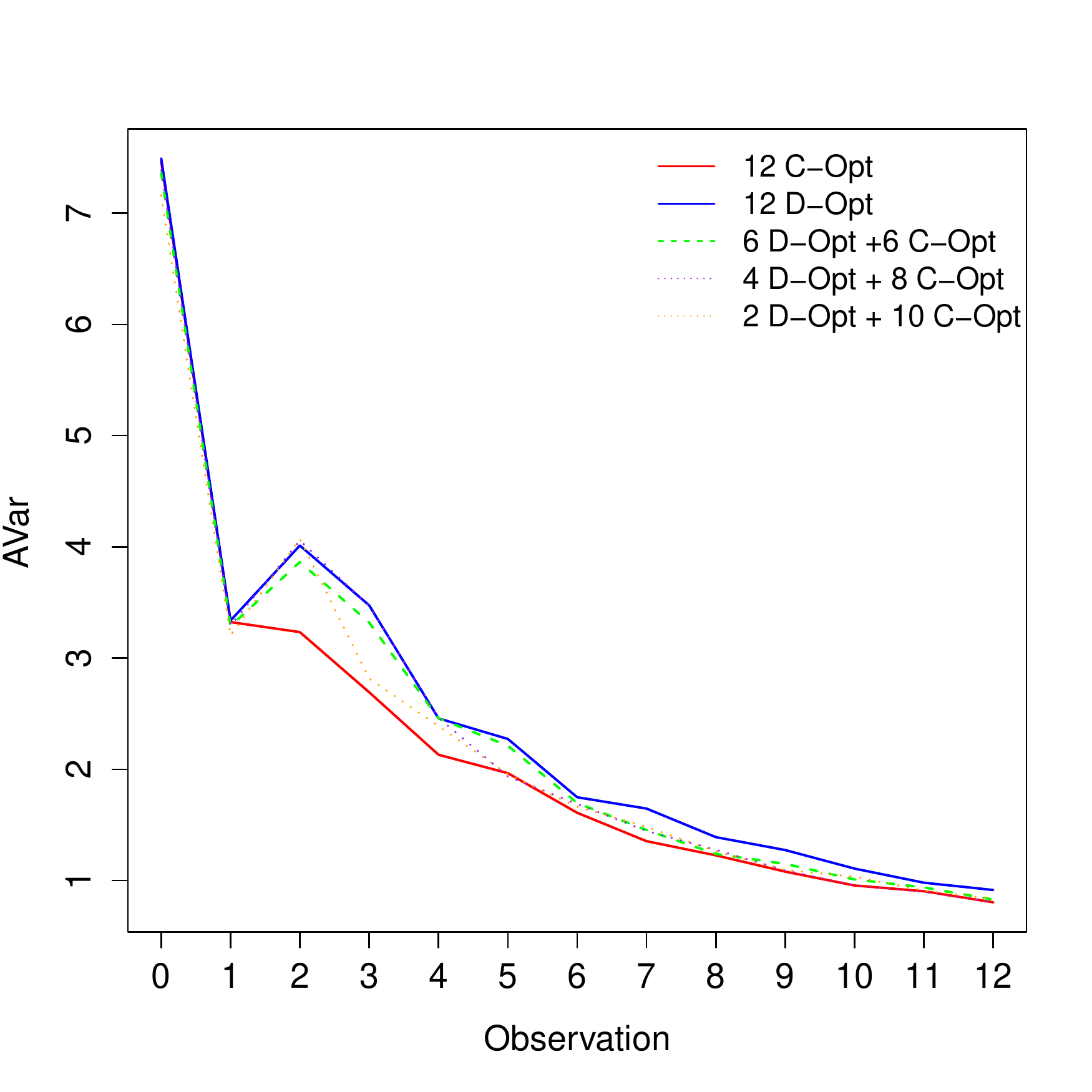}
\end{center}
\caption{Plot of AVar, the estimated asymptotic variance of the estimated quantile life averaged over a range of specified use conditions as in Eqn. (\ref{Eq:wavar2}), for the five designs from Table \ref{tab:fiveSBD}.}\label{fig:AVar_DS1_UCP}
\end{figure}

Given the three selected observations shown in Figure~\ref{fig:DS1} (solid diamonds), we plan to select 12 additional observations using the Bayesian dual objective sequential design. Five different design strategies are compared which are summarized in Table \ref{tab:fiveSBD}. Due to the sampling variation, we simulated 100 test plans based on each strategy and summarize the average performance across the 100 simulations to compare the design performance. Figure \ref{fig:AVar_DS1_UCP} shows the AVar from Eqn. (\ref{Eq:wavar2}) of the 12 sequential runs for the five design strategies. We can see for the first sequential run, all five designs substantially reduce the AVar by almost the same amount. Starting from the 2nd run, the sequential C-optimal design (shown in the red line) achieves the minimum AVar consistently for all remaining runs. The sequential D-optimal design (shown in the blue line) increases the AVar at the 2nd run and then starts to reduce the AVar steadily afterwards. However, it has consistently higher AVar than the sequential C-optimal design since it focuses primarily on obtaining the most precise estimates of the model parameters instead of making the most precise prediction of reliability at the normal use conditions. However, the difference between the two designs becomes noticeably smaller as more sequential runs are observed. The other three designs (c-e) from Table \ref{tab:fiveSBD} achieve AVar values between those offered by the D-optimal and C-optimal designs. Generally, the more D-optimal runs generated at the early stage of the sequential designs, the more closely the design performs compared to the 12 run sequential D-optimal design for those early runs that are generated based on D-optimal criterion.  But for the later runs generated based on the C-optimal criterion, the design usually offers slightly better AVar than the D-optimal design. On the other hand, the fewer D-optimal runs the design has at the early runs, the faster the AVar value is improved and approaching the best AVar offered by the C-optimal design for all the sequential runs.

\begin{figure}
\begin{center}
\includegraphics[width=.5\textwidth]{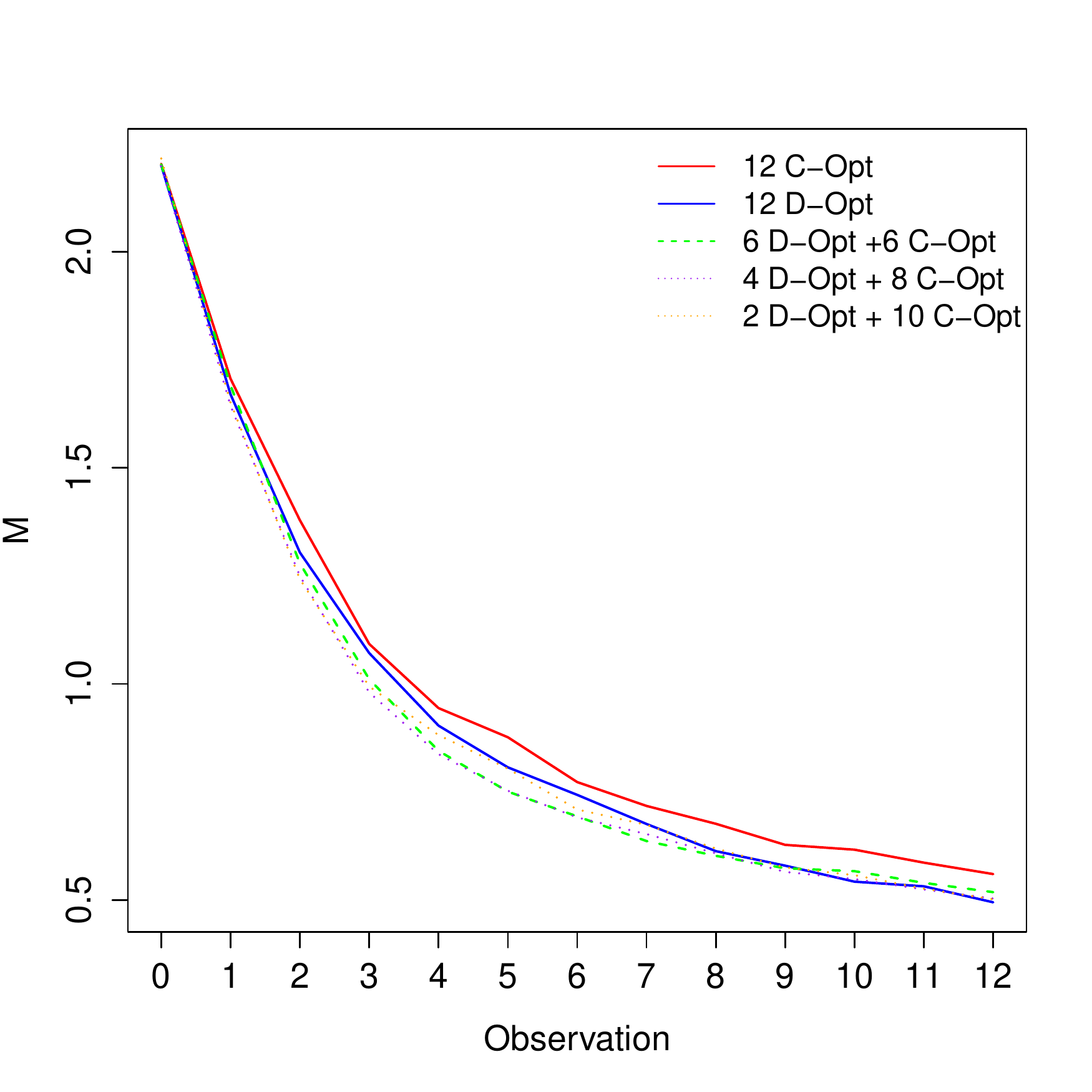}
\end{center}
\caption{Comparison of the $M$ measure, the total relative mean squared error for all three model parameters, for the five designs from Table \ref{tab:fiveSBD}.}\label{fig:M_DS1_UCP}
\end{figure}

In addition, we compare the overall precision of the estimated model parameters among the five design strategies. Particularly, we adopt the $M$ measure for quantifying the relative error of the estimated parameters from Lee et al. \cite{Leeetal2018}. Particularly, define
\begin{equation*}
m\left(\theta_j\right)=\frac{1}{K}\sum_{k=1}^{K}\left(\frac{|\hat{\theta}_{j,k}-\theta_j|}{\theta_j}\right)^2,
\end{equation*}
where $\theta_j$ represents the $j$th parameter in the assumed ALT model, and $\hat{\theta}_{j,k}$ represents the estimated value of $\theta_j$ from the $k$th simulation trial for $k=1,\cdots,K=100$. Hence, $m(\theta_j)$ measures the relative mean squared error of the estimate of the $j$th parameter. Then the $M$ measure, defined as $M=\sum_{j=1}^{3}m\left(\theta_j\right)$, measures the total relative mean squared error of all the parameters, which quantifies the overall precision of the estimated model parameters from the ALT model. The smaller $M$ value indicates more precision of the estimated parameters. Figure \ref{fig:M_DS1_UCP} shows the $M$ measure for five designs shown in Table \ref{tab:fiveSBD}. The sequential C-optimal design has the largest $M$ and hence the least precision of all the estimated parameters for all the 12 sequential runs. This can be expected as the C-optimal design focuses more on obtaining the most precise estimation of the interested quantile at the normal use conditions and less on obtaining the most precise estimation of model parameters. It is interesting that the D-optimal design does not offer the most precise estimation of parameters at the early runs but does catch up later as more runs are implemented. Also, some of the sequential dual objective designs (e.g., design c with 6 D-opt runs followed by 6 C-opt runs) slightly outperform the D-optimal design at early runs (between the 2nd and the 7th sequential runs). This could be resulted from sampling variation and the fact of D-optimality being a better measure of the precision of estimated model parameters for the larger sample cases. As more sequential runs are obtained (8 or more runs), the D-optimal design offers the best precision of estimated model parameters. For our case study, when less than 8 sequential runs are allowed, the dual objective sequential designs offer slightly better precision for both the estimation of model parameters and the prediction at the normal use conditions.

\begin{figure}
\begin{center}
\includegraphics[width=.7\textwidth]{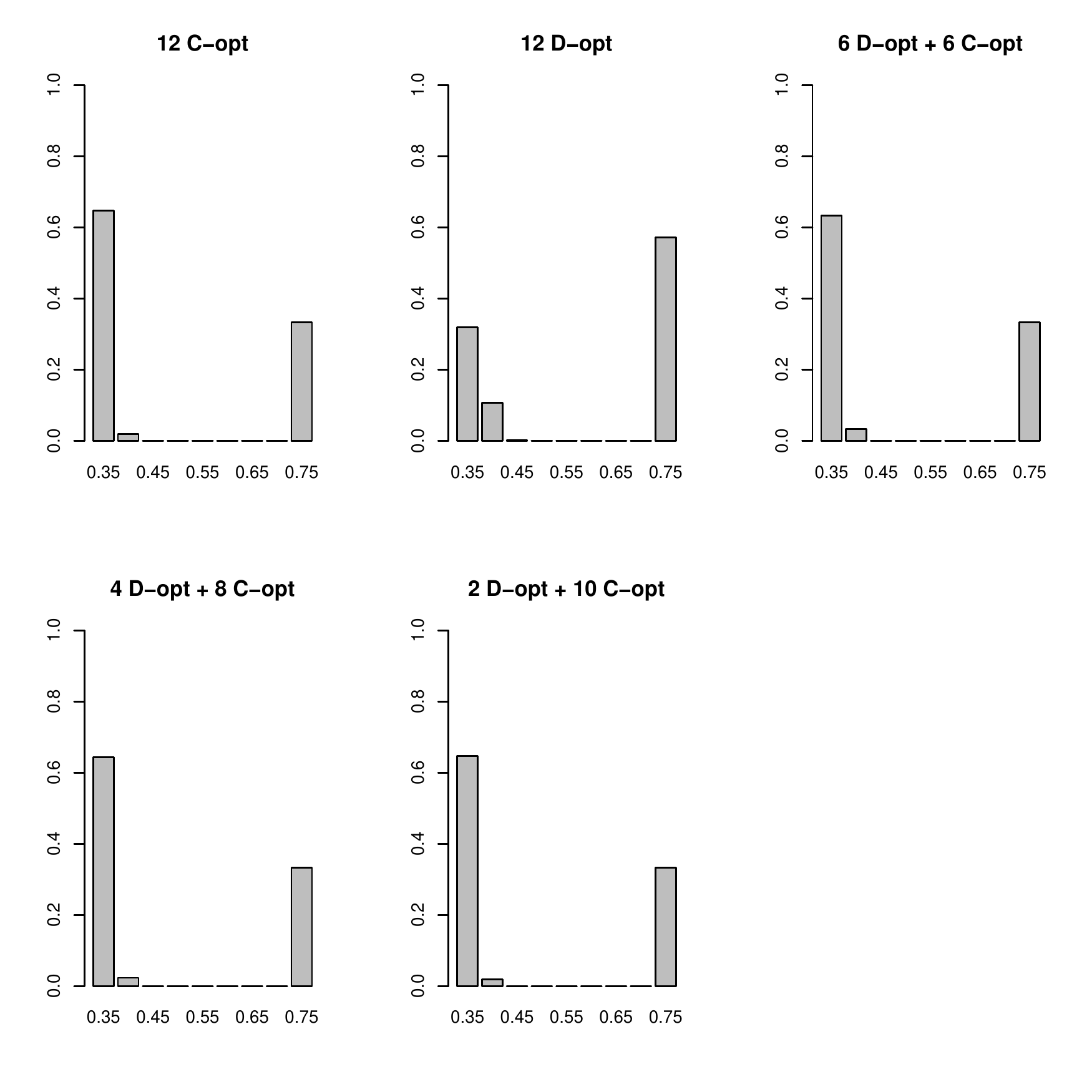}
\end{center}
\caption{Plot of sample size allocation for the five designs from Table 1.}\label{fig:SAlloc_DS1_UCP}
\end{figure}

Figure \ref{fig:SAlloc_DS1_UCP} shows the fraction of sample allocated to different stress levels for the five designs averaged over the 100 simulation trials. A few patterns can be observed prominently. First of all, all the test plans allocate majority of the runs to the extreme stress levels within the range of consideration for $q\in\left[0.35,0.75\right]$. The sequential C-optimal design places about $2/3$ of the total sequential runs to the lowest stress level at $q=0.35$. In contrast, the sequential D-optimal design places about $60\%$ of the total runs to the highest stress level at $q=0.75$. This is intuitive as the C-optimal aims for improving the estimation at the normal use conditions at much lower stress levels than the design region, and hence tends to allocation more runs closer to the region of prediction. On the other hand, the D-optimal design aims to improve the estimated model parameters, which are estimated more precisely when more failures are observed at the higher stress levels assuming the failure mechanism does not change under more stressed conditions. Other sequential plans based on considering dual objectives show slightly more balanced performance between the two optimal designs considering only a single criterion. But they are consistently much closer to the C-optimal design by allocating more runs closer to the normal use considerations. For all the test plans, there are only a small number of runs ($\leq 1\%$) located at the second lowest stress level at $q=0.4$ among the 100 simulation trials.

\begin{figure}
\begin{center}
\includegraphics[width=.8\textwidth]{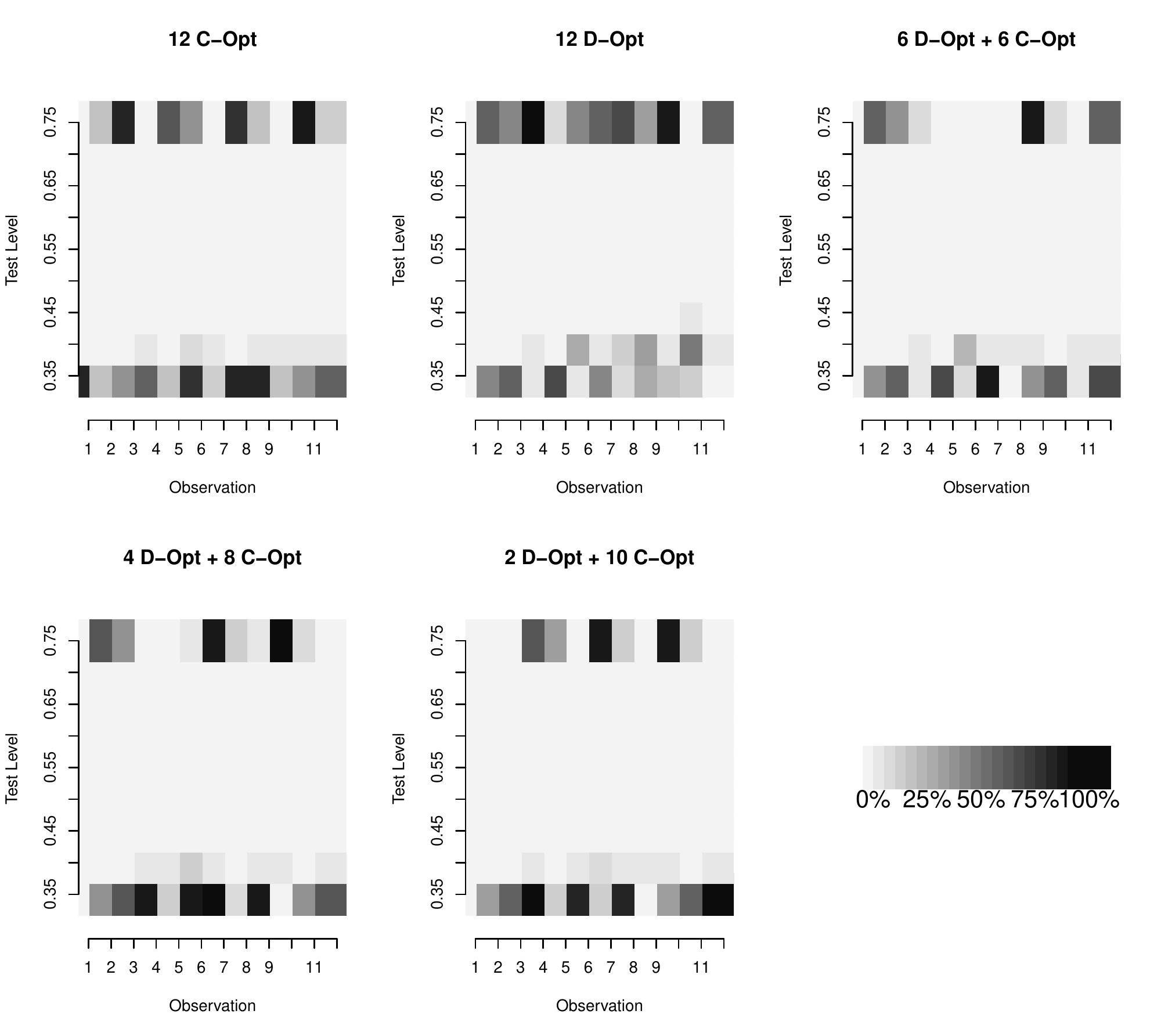}
\end{center}
\caption{Plot of the sample size allocation of the 12 sequential runs for the five designs.}\label{fig:SeqAlloc_DS1_UCP}
\end{figure}

Figure \ref{fig:SeqAlloc_DS1_UCP} displays the faction of sample allocation at each of the sequential runs across the 100 simulation trials for the five designs. The dark-to-light gray shades indicate large to small frequency of sample allocation at the different stress levels. We can see there is a large middle region in each panel (for each design) shown in the lightest gray indicating a rare sample allocation to the middle stress levels between $q=0.4$ to $q=0.7$. By contrasting the C-optimal and D-optimal designs, we can see while the earlier runs are more evenly split between the high and low stress levels, the D-optimal design seems to allocate more runs at the higher stress levels for the later runs, while the C-optimal design tends to have more runs at the lower stress levels at the later stage of the sequential experiment. And this pattern matches our understanding that more tests at the higher stress levels can improve the estimation more efficiently while tests at the lower levels are more helpful for improving the prediction at the normal use conditions. We also did the comparison between the several strategies based on other  data sets, including the Data~Set~2 in Lee et al. \cite{Leeetal2018}, and observed similar patterns. Hence, the results are not shown here.

%%%%%%%%%%%%%%%%%%%%%%%%%%%%%%%%%%%%%%%%%%%%%%%%%%%%%%%%%%%%%%%%%%%%%%%%%%%%%%%%%%%%%%%%%
\section{Discussion and Conclusions}\label{sec:conclusion}
%%%%%%%%%%%%%%%%%%%%%%%%%%%%%%%%%%%%%%%%%%%%%%%%%%%%%%%%%%%%%%%%%%%%%%%%%%%%%%%%%%%%%%%%%
In this paper, we extend the Bayesian sequential test plans by considering dual objectives when selecting the sequential design points. Bayesian sequential tests are particularly useful when there is little prior information on the lifetime distribution and also tight constraints on the total number of test units or the number of units that can be tested simultaneously. In this case, it is helpful to timely update our understanding as more valuable information is collected during the sequential experiment. We choose to use D-optimality to guide the early design selection for quickly improving the precision of estimated ALT models to help the planning strategy focus in the right region and then switch to C-optimality to seek precise prediction when extrapolating into the normal use conditions.

We apply the dual objective Bayesian sequential design to the polymer composites fatigue test problem, in which case there is limited prior knowledge on the new composites materials and the testing requires using an expensive equipment with limited availability and hence there is little chance of testing multiple units at the same time. Our comparison between the dual objective and single objective Bayesian sequential designs showed more robust and balanced performance of the dual objective design. Especially when very limited test units are possibly to be considered, the dual objective designs have shown to offer noticeably more precise estimation of model parameters than the C-optimal design and better prediction across a range of possible normal use conditions than the D-optimal design. As more test units are affordable, the dual objective designs have shown to catch up with the single criterion optimal design more quickly than the optimal design considering only the other criterion and eventually offer near-optimal performance on both the estimation and prediction.

The implementation of the method is convenient and straightforward with the ``SeqBayesDesign" R package \cite{packageSeq} for the polymer composites fatigue test example and other applications using similar ALT models. However, the proposed method is very general to be applied to broad applications with different lifetime distributions, ALT models, or even different sequential testing strategies involving blocking or other constraints. The method should also adapt for higher dimensional problems with more than one accelerating factors.

%%%%%%%%%%%%%%%%%%%%%%%%%%%%%%%%%%%%%%%%%%%%%%%%%%%%%%%%%%%%%%%%%%%%%%%%%%%%%%%%%%%%%%%%%%%%%%%%%%%%%%%%%%%%%%%%%
\section*{Acknowledgments}
%%%%%%%%%%%%%%%%%%%%%%%%%%%%%%%%%%%%%%%%%%%%%%%%%%%%%%%%%%%%%%%%%%%%%%%%%%%%%%%%%%%%%%%%%%%%%%%%%%%%%%%%%%%%%%%%%
The authors thank the editors and one referee who provided comments that helped us improve this paper.  The authors acknowledge Advanced Research Computing at Virginia Tech for providing computational resources. The work by Hong was partially supported by the National Science Foundation under Grant CNS-1565314 to Virginia Tech.

%%%%%%%%%%%%%%%%%%%%%%%%%%%%%%%%%%%%%%%%%%%%%%%%%%%%%%%%%%%%%%%%%%%%%%%%%%%%%%%%%%%%%%%%%


\end{document}